\magnification=1400

\centerline{\bf Particle Physics in The United States}
\centerline{\bf A Personal View}
\centerline{\bf Sheldon Lee Glashow}
\centerline{\bf Boston University}
\centerline{\bf \tt slg@bu.edu}
\bigskip

\noindent ABSTRACT: {\sl I present my views on the future of 
America's program in particle physics. I discuss a variety
of experimental initiatives  that do have  the potential to make
transformative impacts on our discipline and  should be included
in our program, as well as others that do not and should not.}
 
\bigskip

This is a notable year in the history of our discipline.\footnote{}
{Research supported in part by the US Department of Energy under Grant
Number DE-SC0010025.}  For six decades America  led the world in
the pursuit of particle physics at the highest energies: at the
Brookhaven Cosmotron (commissioned in 1953), the Berkeley Bevatron one
year later and at their more powerful successors up to the Fermilab
Tevatron Collider. Sadly, with the cancellation of the SSC project
twenty years ago, last year's shutdown of the Tevatron Collider and my
country's failure to formulate any plan to regain the initiative, this
heroic era of American leadership in high-energy physics has ended.  For
the forseeable future the high-energy frontier will be explored on
foreign soil and under foreign control.  \medskip

By a remarkable coincidence 2013 also marks the opening of a new and
exciting chapter for particle physics  in which all of the ingredients of
our standard model have been seen and accounted for. Of the final
three, top quarks and tau neutrinos were found at Fermilab in 1995 and
2000 respectively, while the Higgs boson held out until last year
when it was first produced and detected at CERN's Large Hadron
Collider.  \medskip

What will be the future of high-energy physics and  what role can the
US play?  In the near term (a few decades?) the LHC
{\bf is} the high-energy frontier and I hope my country (and my
university) will continue its active and effective engagement at
CERN.  Although the world's next great  collider is unlikely to be built
in the US,  I hope that we will be eager participants in any sensible
future multinational efforts. But can there be a fruitful domestic
program in experimental particle physics as well?

\vfill\eject

These  questions must be considered by all stakeholders
in the American particle physics enterprise so as to maximize the
scientific impact of scarce resources.  Here I offer my own 
thoughts,  mostly concerning high precision 
experiments designed to detect tiny effects
or rare and as yet unobserved  processes.
Demands for ever more accurate measurements of physical
quantities have a long history:

{\smallskip\narrower\narrower\noindent{\it There is nothing new to be
  discovered in physics.\hfill\break
 All that remains is more and more precise
  measurement,..}\hfill\break
$\phantom{sheldonsheldonsheldon}$ William Thompson (1900)\smallskip}

{\narrower\narrower\noindent{\it The whole history of
physics proves that a new discovery is quite likely lurking at the
next decimal place...\quad}\hfill\break
$\phantom{sheldonsheldonsheldon}$ F.K. Richtmeyer (1931)\smallskip}

\noindent Early 20th century emphases on the virtues of precision {\it
per se\/} were neither wise nor productive. Most soon-to-arrive
seminal advances had little to do with increased accuracy: Think of
discoveries such as parity and CP violation, J/Psi particles and tau
leptons, neutrino masses and their  oscillations, neutral currents and the
Higgs boson. These transformative revelations resulted from looking
carefully where few had looked before.  And yet, several exciting
developments did arise from the simple pursuit of precision, among
them observations of Uranus' orbit enabling Neptune's discovery,
measurements of gas densities leading to the discovery of inert gases
and spectroscopic observations driving the development of QED. I dare
to imagine that it may happen again!

\medskip

Many significant findings may still be ``lurking at the next decimal
place.''  But beware! Most leaps to the next decimal place accomplish
just that, but nothing more.  The proton magnetic moment is known to
ten ppb. Increasing that precision to one ppb seems pointless.
Sensibly, nobody has proposed doing so. Similarly, the MuLAN
experiment achieved its goal of measuring the muon lifetime to within
one ppm and thereby the Fermi constant to 0.5~ppm.  But of the three
conventional quantities parameterizing the electroweak theory,
$\alpha$, $G_F$ and $M_W$, the last is known to only 23 ppm.  MuLAN's
heroic result may be of metrological interest but has not and could
not have revealed a secret of nature.  This precision experiment,
however effectively it was carried out, was certainly not
transformative.  \medskip

Consider the next-generation of muon $g-2$ experiments. Some years ago
 Brookhaven determined $g_\mu-2$ to the remarkable precision of 0.5
 ppm, a result differing by over 3$\sigma$ from its theoretical value,
 whose uncertainty is also said to be $\sim\! 0.5\;$ppm.  Although
 this discrepancy is intriguing, estimates of the theory error
 depend on model-dependent calculations of the hadronic light-by-light
 contribution. Future $g-2$ experiments planned for Fermilab and J-PARC
 can  reduce the experimental error substantially, possibly yielding a
 much more significant discrepancy, which would become decisive
 evidence for new physics if and when lattice calculations yield a robust
 bound to the light-by-light contribution. I now turn to other pursuits
 of high precision which  may make immediate and transformative
 discoveries.  \medskip

$\bullet$ {\it Dark Mass \&\ Dark Energy:} These are two of the most
puzzling features of the universe.  Efforts to clarify them certainly
deserve support and encouragement. The study of dark energy lies
exclusively at the cosmic frontier which, like the high-energy
frontier, I do not consider herein.  The study of dark matter has a
cosmic component (searches for decay or annihillation), a high-energy
component (production and detection) and a terrestrial component,
where many experiments have tried and failed to detect the passage of
cosmic dark matter particles through matter.  The need for greater
precision is evident. Direct detection of dark matter would imply that
it possesses other than gravitational interactions, thus
opening many new avenues for research.

\medskip
$\bullet$ {\it Testing Global Symmetries:} How exciting it would be 
to discover violations of either of our two surviving exact global
conservation laws: lepton number $\cal L$ and baryon number
$\cal B\,$! The relevant processes are single nucleon decay ($\Delta {\cal
B}=1,\ \Delta {\cal L}=\pm 1$), neutron-antineutron oscillation or
dinucleon decay 
($\Delta{\cal B}=2,\ \Delta{\cal L}=0$) and neutrinoless double beta
decay ($\Delta {\cal B}=0,\ \Delta{\cal L}=2$). All have been looked for,
none have been seen. More sensitive searches should be launched because
the confirmed violation of any of these symmetries  would be  profoundly
important.  \medskip

For decades nucleon decay has been a target for experimenters: at IMB,
 Kamiokande, Super-K and elsewhere. Very strong limits were set
 on many $\Delta {\cal B}=1$ and $2$ decay
 modes. Significantly improved searches would be very costly
 in both time and money. They may be done eventually
 (perhaps at Hyper-K) but cannot be part of the domestic US program at
 this time.  \medskip

Lower bounds of about three years have been set on the free
neutron--antineutron oscillation period. The construction of a new
(and costly) slow neutron source may allow the bound to be increased
upwards of a thousandfold. Such a deep plunge into the unknown
must be attempted, even though I can think of no plausible argument
that either suggests or forbids a positive result.  \medskip

Several searches for neutrinoless double beta decay were successfully
 completed, leading only to stronger lower bounds on lifetimes.
 Other experiments are planned, proposed or proceeding. Calorimetric
 searches also have been proposed 
for a related  process:
neutrinoless double electron capture.
 Both modes of neutrinoless double beta decay  are allowed if neutrino masses are
lepton-number violating ({\it i.e.,} Majorana rather than
 Dirac). Their
 rates are 
 controlled by $|m_{\beta\beta}|$, the magnitude of the $e$-$e$ entry
 of the neutrino mass matrix whose current upper bound is $\sim\!
 1\;$eV.  Perhaps the US should organize and lead a far more aggressive
 and better funded assault to shrink the bound as much as possible...
 or better, to prove that lepton number is not conserved.

\medskip

$\bullet$ {\it Testing Flavor Symmetries with Muons:} I focus on 
these  three flavor-changing muon decay modes: radiative decay
($\mu\rightarrow e+\gamma$), 3-e decay ($\mu\rightarrow
e+e+\overline{e}$) and orbital conversion ($\mu+ {\cal N}\rightarrow e+
{\cal N}$) where $\cal N$ denotes an atomic nucleus to which the muon
is bound.  These decay modes conserve $\cal L$ and are technically
allowed via  the flavor-violation responsible for neutrino oscillations.
 Because their standard-model branching ratios are far
too tiny  for possible detection,  observation of any
 mode  would be certain evidence of new physics.  That's
what makes such sensitive searches potentially
transformative.  \medskip

Whatever  enables  one muon mode enables the others as well, but often
with considerable suppression. If new physics primarily generates
flavor-changing lepton magnetic moments, $\mu\rightarrow e+\gamma$
would dominate the other modes by factors exceeding 200. Likewise,
doubly-charged dileptons could strongly favor the 3-e mode while
leptoquarks could strongly favor orbital conversion.  Thus an improved
search for one mode does not obviate the need to search for the
others. Current bounds on all three modes  should be
improved as much as possible.  (Flavor-violating decays of
tau leptons, such as $\tau\rightarrow \mu + {\rm hadrons}$ or
$\tau\rightarrow \mu+\mu+\overline{e}$, may be worth further
constraining, but these experiments cannot approach the sensitivity
accessible to the three golden muon modes.)\medskip

$\bullet$ {\it Electric Dipole Moments:} Standard-model CP violation
seems insufficient to explain the observed baryon asymmetry of the
universe. The required new physics, lying  at inaccessably
high  energy, may 
 induce tiny but observable electric dipole moments of leptons and
nucleons, for which current upper limits are:
$$d_e<7\times 10^{-28},\quad d_\mu< 1.8\times 10^{-19},\quad d_n<
3\times 10^{-26}$$
in units of $e\,$cm.
These bounds are many powers of ten too weak to reveal
standard-model effects. However,  for electrons and neutrons 
they are tantalizingly close to electric
dipole moments expected in  several theoretical schemes..  Future
experiments should  achieve a precision of $10^{-29}\,e$cm  for
both neutrons and electrons.
Both searches are needed because the results would complement one
another and further illuminate the nature of the underlying
new physics.
Searches for the muon electric dipole moment are far less likely to
attain the necessary precision.

\medskip

$\bullet$ {\it Neutrino Physics:} First, a brief review: Three active
neutrino states, each in a weak doublet, suffice to describe
all confirmed  neutrino phenomena. Their oscillations involve
six parameters: two squared-mass differences ($\Delta_s\equiv
m_2^2-m_1^2$ and $\Delta_a\equiv |m_3^2-m_2^2|$), three mixing angles
($\theta_{12},\, \theta_{23}$ and $\theta_{13}$) and the CP-violating
angle  $\delta$.  All of these parameters except $\delta$ have been
measured to an accuracy of 20\% or better.  The overall neutrino mass
scale is severely constrained, while    $\delta$ is not
constrained at all.
\medskip

  I shall mention   three  potentially transformative
experiments (and one astrophysical challenge)  that might
be  parts of  the domestic US
program: one involves CP violation, two  concern neutrino
masses and one  has to do with hypothetical `sterile'
neutrinos. Certain other experiments, such as increasingly precise
measurements of $\theta_{ij}$ and $\Delta_{s,a}$, may be worth doing
{\it en passant\/} but are  unlikely to
yield surprising, illuminating or even particularly useful results.
\medskip

Is observable CP violation  confined to
hadrons?  I would assign very high  priority to experiments that could
demonstrate the existence of  CP violating effects  in the neutrino sector. The
accuracy with which  oscillation parameters are already
known surely suffices for the design of an experiment that can
accomplish this  goal.  
Two additional  challenges relate specifically 
to neutrino masses.
Firstly, we must establish  their overall mass scale. A partial answer may
emerge from beta-decay endpoint measurements  or  from the detection of
neutrinoless double beta decay. More likely,  the sum of 
the  neutrino masses will be determined 
at the cosmic frontier by precise astrophysical
observations such as  have already yielded impressive  upper bounds. 
\medskip

The other important mass-related issue is the binary choice between two
orderings of  neutrino masses.  For the normal mass hierarchy, the
dominant constituents of electron neutrinos, $\nu_1$ and $\nu_2$, are
less massive than $\nu_3$, whereas for the inverted hierarchy
$m_2>m_1>m_3\ge 0$.  Several experiments have been proposed to answer this
question, thereby providing  extremely helpful hint to model-building
theorists. The result is also relevant to the design of experiments to search
for CP violation in  neutrino oscillations.  \medskip

My last  neutrino initiative is related to recent
determinations of the reactor angle $\theta_{13}$. 
The observed {\bf disappearance} of reactor-produced
electron neutrinos is presumed due to their oscillation into muon or
tau neutrinos, an effect controlled by $\theta_{13}$. 
But a quite different  explanation is conceivable:  that  reactor neutrinos
oscillate into light sterile neutrinos.
Experimenters can rule out this possibility  by  observing the {\bf appearance}
of electron neutrinos in an accelerator-produced beam of muon neutrinos.
 I would not be surprised if such an experiment succeeded in
confirming the orthodox view, but I would be astounded and delighted if
it did not.\medskip

$\bullet$ {\it Conclusion:} In structuring America's plan for future
 research in elementary particle physics we must avoid the pursuit of
 accuracy for its own sake. A precision experiment is justified if it
 can reveal a flaw in our theory or observe a previously unseen
 phenomenon, not simply because the experiment happens to be feasible
 or the quantity being measured is a so-called ``fundamental constant
 of nature.'' Surely much remains to be discovered about elementary
 particles and the domestic American experimental program can have a
 richly rewarding and most exciting future.  I have confined myself here
 almost exclusively to  promising research directions lying away
 from the `high-energy' and `cosmic' frontiers where many other
 delights surely await us.
Even so, I must surely have forgotten to mention or failed to imagine
many exciting challenges at the high-luminosity frontier.

\bigskip

I am pleased to thank 
Barry Barish, Andrew Cohen, Alvaro De R\'ujula, John Iliopoulos and
Kenneth Lane for their advice and encouagement.

\bye

\bigskip
I 
three  potentially transformative neutrino challenges.

 The other is the search for CP violation in the context of
neutrino oscillations. Now that all three mixing angles (especially
the smaller `reactor angle' $\theta_{13}$) are roughly known,
 we know that CP-violating effects may be measurable. CP violation
  remains puzzling  because we need much more of it to explain the
 cosmological baryon asymmetry. Thus, it is important to discover
 whether the hypothetical source of CP violation percolates down to
 the neutrino arena.

\bye